\documentclass[12pt]{article}
\usepackage{epsfig}
\usepackage{mathtools}

\def\beq{\begin{equation}}
\def\eeq#1{\label{#1}\end{equation}}
\def\eeqn{\end{equation}}

\def\beqa{\begin{eqnarray}}
\def\eeqa#1{\label{#1}\end{eqnarray}}
\def\eeqan{\end{eqnarray}}

\let\bar=\overbar

\def\Dslash{\not{\hbox{\kern-4pt $D$}}}
\def\dslash{\not{\hbox{\kern-2pt $\del$}}}

\def\msb{{\bar{\ssstyle M \kern -1pt S}}}

\def\Title#1{\begin{center} {\Large {\bf #1} } \end{center}}

\begin{document}

\Title{Quantum Confinement Effects for Semiconductor Clusters in the Molecular Regime}

\bigskip\bigskip

%+\addtocontents{toc}{{\it D. Reggiano}}
%+\label{ReggianoStart}

\begin{raggedright}  

{\it John Zhang\index{Zhang, J.}\\
Altamont Research\\
Altamont, NY12009}
\bigskip\bigskip
\end{raggedright}

{\bf Abstract:}  Based on the observed absorption spectral band shifts, the growth process of the semiconductor clusters was divided into two phenomenological regimes: The "molecular regime" that is associated with the band blue shift as the size of cluster increases and the "crystallite regime" that is associated with the band red shift as the size of cluster increases. We show that in the molecular regime, the band blue shift associated with cluster growth can be understood by a model that assume electrons are confined to a spherical potential well and the clusters are made of some basic units. A formula is given for the lowest excited electronic state energy. This expression contains an electron-hole-pair (EHP)  delocalization constant $ \zeta $ as an adjustable parameter which, however, can be anchored to a definite value through the known transition energy at the spectra turn-around point. The stability of clusters is characterized by a function $\Delta _2 ( N ) $  that can be calculated by the eigenvalues of the Hamiltonian of the model.

\section{Introduction}

Significant motivation for studying clusters comes from both basic and applied science. The transitions involved in grouping atoms to form extended solids are not understood in detail; hence the study of the relationship between the properties of atoms, molecules, clusters, solid surfaces and bulk solids has attracted considerable attentions\index{Cohen}~\cite{Cohen}.
Recently, size regimes in the evolution of semiconductor spectroscopic properties have been labeled with increasing size as molecular, quantum dot, polariton, and finally bulk semiconductor species by Brus et al\index{Brus}~\cite{Brus}, Broadly speaking, these species represent the evolution of molecular to unit-cell structures, discrete electronic states to continuos bands, and weak dipole scattering to strong macroscopic, polariton electromagnetic scattering~\cite{Brus1}.

	In quantum dot regime, various semiconductor quantum dots have been shown that their optical properties are sensitive to size~\cite{Brus2,Henglein,Kuczynski,Wang,Rossetti,Vossmeyer}. Tremendous experiments gave the reproducible results on such properties as the electronic absorption shifting to lower energy with increasing size of the clusters (red shift). This is usually recognized as a quantum confinement effect.
      
      To explain the three-dimensional quantum confinement, the quantum mechanical problem of the motion of a particle in a box has been improved during the last few years by adding more and more complicated expressions to the Hamiltonian~\cite{Woggon,Banyai},taking into account, for example, electron-hole Coulomb interaction, a more complicated valence-band structure or non parabolic bands; starting with one-electron-hole-pair(1EHP) states and going to two pair (2EHP) and many-particle states etc~\cite{Woggon}.
      
      In the 1EHP state, taking into account for the electron-hole Coulomb interaction, the Hamilton of the dot can be written as following:

\begin{equation}
\hat{H}=\left.-\frac{\hbar^2\nabla_e^2}{2m_e}-\frac{\hbar^2\nabla_h^2}{2m_h}-\frac{e^2}{\varepsilon|r_e-r_h|}\right.+V_e(\overrightarrow{r_e})+V_h(\overrightarrow{r_h})
\label{equ:a}
\end{equation}

Here $m_e (m_h)$ is the effective mass of the electron (hole), $\varepsilon$  is the semiconductor dielectric constant. This Hamiltonian has been treated by perturbation theory~\cite{Brus3,Brus4},
variational calculations~\cite{Schmidt,Nair,Kayanuma,Ekimov,Thoai,Takagahara}, Monte-Carlo-technique~\cite{Pollock}  and matrix diagonalization methods~\cite{Hu,Park}. Treating the problem in perturbation theory, Brus got the lowest excited state energy as following~\cite{Brus3,Brus4}:

\begin{equation}
\ E=\frac{h^2 n^2}{8R^2}\left[\frac{1}{m_e}+\frac{1}{m_h}\right]-\frac{1.8 e^2}{\varepsilon R}+E_g
\label{equ:b}
\end{equation}

Where n is an integer number, h is the Planck's constant, e is the elementary charge, $m_e (m_h)$ is the effective mass of the electron (hole), R is the cluster radius,  $\varepsilon$ is the bulk dielectric constant, and $E_g$ is the bulk band gap. This famous formula shows a fact that the absorption energy of semiconductor quantum dot should shift to lower energy as the size of the cluster increases (red shift). This kind of quantum confinement effect has been confirmed by a lot of experiments both for direct~\cite{Henglein,Kuczynski,Wang,Rossetti} and indirect~\cite{Chen,Chen1,Ehrlich}semiconductor quantum dots.

\begin{figure}[htb]
\begin{center}
\epsfig{file=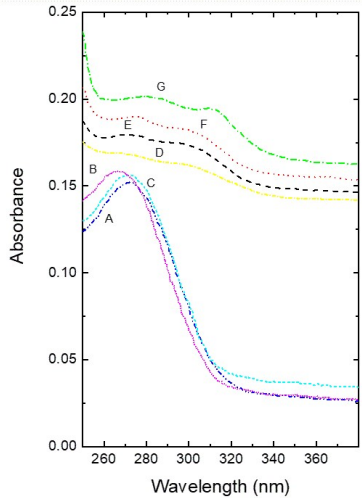,height=3.0in}
\caption{Curves A-C: Absorption spectra of AgBr quantum Dots produced from Ag+ (trapped in DOPC vesicles) and Br- ions (in the bulk solution). Using electroporation, the growth of the cluster is restricted to the “molecular regime”: A-0.1h, B-5h, and C-11h after application of the high-voltage square pulse. The corresponding shifts of absorption maxima are : 274nm to 269nm and then to 273nm. Curves D-G: Clusters of the “crystallite regime” are formed (D-0.1h, E-1h, F-3h, and G-20 h after addition of Brij-35, without electroporation) in a system where both the Ag+ and Br- ions are individually trapped in different vesicles. See ~\cite{Correa}}
\label{fig:figure1}
\end{center}
\end{figure}

Compare to the huge amount of work done in the quantum dot regime, few work has been done in the molecular regime. Recently, H.Zhang and M.Mostafavi used pulse radiolysis to study the formation of AgBr molecule and the $(AgBr)_2$ dimer in solution~\cite{Zhang}. They found that a short time (400ns) after the nanosecond pulse, the kinetics of AgBr formation is observed at 295nm. The monomer signal decays to give the dimer $(AgBr)_2$ . The dimer spectrum is shift to the blue and the maximum is at 285nm. Most recently, Schelly et al observed the absorption spectrum of AgBr clusters in the molecular regime first shift to blue from 274nm to 269nm and then turn round back to 273nm as the size of clusters increases. The subsequent further growth of the cluster is manifested in a red-shift of two bands split from the original one by spin orbital interaction as shown in Figure~\ref{fig:figure1}~\cite{Correa}.Similar spectra of the growth process have also been observed by Meisel et al in the case of AgI cluster~\cite{Schmidt1}and by Wang et al in the case of PbS cluster~\cite{Wang1}.

      Based on the above kinds of observations of size dependent absorption spectra, the growth process of the semiconductor clusters can be divided into two phenomenological regimes: The "molecular regime" that is associated with the absorption band energy shift to higher energy as the size of cluster increases (blue shift) and the "crystallite regime" that is associated with the absorption band energy shift to lower energy as the size of cluster increases (red shift). We show that in the molecular regime, the band blue shift associated with cluster growth can also be understood by the model that assume electrons are confined to a spherical potential well and the clusters are made of some basic units. A formula is given for the lowest excited electronic state energy. This expression contains an electron delocalization constant   as an adjustable parameter which, however, can be anchored to a definite value through the known transition energy at the spectra turn-around point. The stability of clusters is characterized by a function   that can be calculated by the eigenvalues of the Hamiltonian of the model.

\section{Size Dependent Absorption Spectrum }
\subsection{Particle in the Spherical Quantum Well Model}
Consider the electrons are confined in a spherical, infinite potential well, each electron has an effective mass $ \mu $~\cite{Xia}.The Hamilton is given by:

\begin{equation}
\hat{H}=\left.\frac{\hbar^2}{2m_\mu}\right.\nabla_\mu ^2-V_\mu (r)
\label{equ:c}
\end{equation}

With the confinement potential:

\begin{equation}
\ V_\mu(r) = \begin{cases} 
   0 & \text{for } r \leq R \\
   \infty      & \text{for } r> R
  \end{cases}
\label{equ:d}
\end{equation}

In the spherical coordinate system, the stationary schrödinger equation of the particles in the above confinement potential is:

\begin{equation}
\left.\frac{1}{r}\bullet\frac{\partial}{\partial r}(r^2\frac{\partial \psi}{\partial r})\right.+\left.\frac{1}{r^2\sin \theta}\bullet\frac{\partial}{\partial \theta}(\sin \theta\frac{\partial \psi}{\partial \theta})\right.+\left.\frac{1}{r^2\sin \theta}\bullet\frac{\partial^2 \psi}{\partial \varphi ^2}\right.+\left.\frac{2\mu E\psi}{\hbar ^2}\right.=0
\label{equ:e}
\end{equation}

Equation~\ref{equ:e} \index{equ:e} has solutions:

\begin{equation}
\psi_{nlm}(r,\theta,\varphi) =Y_{lm}\bullet N_{nl}\bullet j_l\left(\sqrt{\frac{2\mu E}{\hbar^2}}r\right)
\label{equ:f}
\end{equation}

With $-l\leq m \leq l, l=0,1,2,3, \dots, n=1,2,3,\dots $. Here $ j_l$  are the sphere Bessel functions~\cite{Abramowitz},  $N_{nl} $ is the normalization constant and $ Y_{lm}$  are the spherical harmonics. The energy eigenvalues $E_{nl} $  follow from the requirement that the wave function has to vanish at $r=R$  :

\begin{equation}
 j_l\left(\sqrt{\frac{2\mu E}{\hbar^2}}r\right)|_{r=R} =0
\label{equ:g}
\end{equation}

and $E_{nl}$  is determined by the zeros of $ j_l(\chi)$  as:

\begin{equation}
 \ E_{nl}=\left. \frac{\hbar^2}{2\mu R^2}\right.\bullet\chi_{nl}^2 
\label{equ:h}
\end{equation}

  $\chi_{nl}$  is the $n^{th} $ zero  of the spherical Bessel function of the order $ l $ ~\cite{Abramowitz},  $\mu$ is the effective mass of the electron and R is the radius of the nanocrystals. Labeling the quantum numbers  $l=0,1,2,3,\dots\dots$  with the letters  $s,p,d,f,\dots\dots$  , the first roots are  $\chi_{1s}=3.142,\chi_{1p}=4.493,\chi_{1d}=5.763 $ etc. These values have been listed in Table~\ref{tab:shell}.

\begin{table}[htb]
\begin{center}
\begin{tabular}{l|ccccc} \hline 
Shell & $ \chi_{nl}$ &$ l $ & Degeneracy & $N_{closed}$& $ N_{total}$\\ \hline 
1s &  3.142   & 0   & 1 &  2&  2\\
1p &  4.493   & 1   & 3 &  6&  8\\
1d &  5.763   & 2   & 5 & 10& 18\\
2s &  6.283   & 0   & 1 &  2& 20\\
1f &  6.988   & 3   & 7 & 14& 34\\
2p &  7.725   & 1   & 3 &  6& 40\\
1g &  8.182   & 4   & 9 & 18& 58\\
2d &  9.095   & 2   & 5 & 10& 68\\
1h &  9.356   & 5   & 11& 22& 90\\
3s &  9.425   & 0   & 1 &  2& 92\\
2f &  10.417  & 3   & 7 & 14&106\\
1i &  10.513  & 6   & 13& 26&132\\
3p &  10.904  & 1   & 3 &  6&138\\
2g &  11.705  & 4   & 9 & 18&156\\
3d &  12.323  & 2   & 5 & 10&166\\
4s &  12.566  & 0   & 1 &  2&168\\
2h &  12.966  & 5   & 11& 22&190\\
3f &  13.698  & 3   & 7 & 14&104\\ \hline
\end{tabular}
\caption{Shell structure data for particles in the spherical quantum well model. are electron numbers for a closed shell and  is the total electron number in a closed shell.}
\label{tab:shell}
\end{center}
\end{table}

\subsection{Size Dependent Absorption Spectrum}
Consider the $ s $  state absorption spectrum, where $ l=0 $ , we have:

\begin{equation}
 \ j_0=\left. \frac{\sin \chi}{\chi}\right.
\label{equ:i}
\end{equation}

From equation~\ref{equ:g} \index{equ:g} we can obtain: $\chi_{0l}=n\pi $ , $n=0,1,2,3, \dots\dots $ , so the electronic energy state are:

\begin{equation}
 \ E_n=\left. \frac{n^2 h^2}{8\mu R^2}\right.
\label{equ:j}
\end{equation}

Assume $ n $  is the quantum number of highest occupied orbital, the lowest absorption energy for the semiconductor cluster is:

\begin{equation}
 \ \Delta E= E_{n+1}-E_n \approx  \left. \frac{n^2 h^2}{4\mu R^2}\right.  (\text{for } n \gg  1 )
\label{equ:k}
\end{equation}

It is reasonable to consider semiconductor clusters are made of some basic units~\cite{Alivisatos,Chen2}. Each unit has $ N $  electrons and volume $ V $ , $ N=N_0 \zeta$  . Here  $\zeta$ is the delocalization constant. $ \zeta=1 $   when the electrons are completely delocalized and $\zeta=0$  when the electrons are completely localized.  $N_0$ is the number of electrons in each basic unit when the electrons are completely delocalized. The total number of electrons can be calculated as:

\begin{equation}
 \ n_{tot} = \left. \frac{4/3 \pi R^3}{V}\right. \bullet N=\left. \frac{4 \pi R^3}{3V}\right.\bullet N_0\zeta
\label{equ:l}
\end{equation}

Consider each orbital can be occupied by two electrons, we have:

\begin{equation}
 \ n= \frac{ n_{tot}}{2} =\left. \frac{2\pi R^3}{3V}\right.\bullet N_0\zeta
\label{equ:m}
\end{equation}

Substitute equation ~\ref{equ:m} to equation ~\ref{equ:k}, we get the absorption energy as:

\begin{equation}
 \ \Delta E= \left. \frac{\pi h^2 N_0}{6\mu V}\right. \bullet \zeta R
\label{equ:n}
\end{equation}

\begin{figure}[htb]
\begin{center}
\epsfig{file=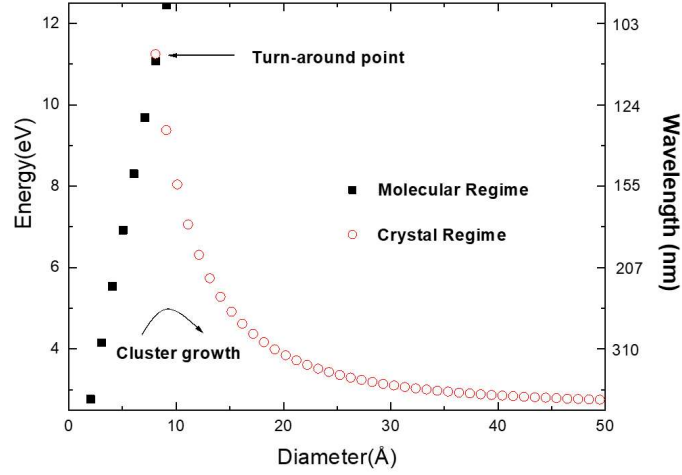,height=3in}
\caption{The correlation of the absorption energy at peak wavelength versus the diameter of silver bromide clusters both for the molecular regime(dot)and the crystal regime(circle).}
\label{fig:figure2}
\end{center}
\end{figure}

\begin{figure}[htb]
\begin{center}
\epsfig{file=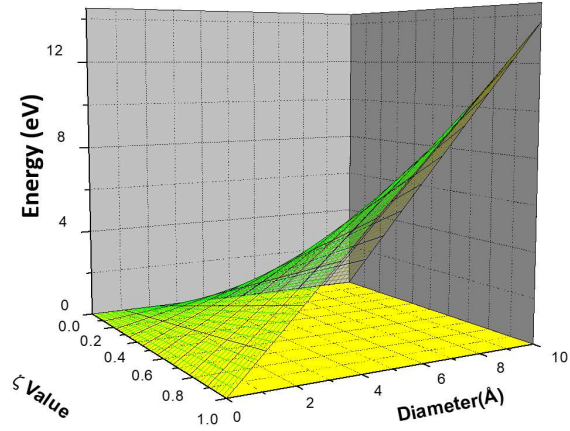,height=3in}
\caption{Absorption energy of clusters in the molecular regime as a function of the delocalization constant   and diameter of the cluster.}
\label{fig:figure3}
\end{center}
\end{figure}

From equation ~\ref{equ:n}\index{equ:n}, we can see that the absorption energy is proportional to the radius R, so as the size of cluster increases, the absorption energy will shift to higher (blue shift). This is the characteristic band shift of clusters in the molecular regime shown in Figure~\ref{fig:figure1}. Based on the observed spectral band shifts, one can estimate the size of semiconductor clusters from molecular to crystal form from Equation  ~\ref{equ:b} and  ~\ref{equ:n}. Figure~\ref{fig:figure2} shows the correlation of the absorption energy at peak wavelength versus the diameter of silver bromide clusters. In the crystal regime, the curve with circles is obtained according to Brus formula equation  ~\ref{equ:b} by using the effective mass of electron $ m_e=0.286 $~\cite{Spoonhower,Hodby}  the effective mass of hole $ m_h=1.096 $~\cite{Spoonhower}, the optical frequency dielectric ${\varepsilon=6.103(400nm)}$, and the band gap of 2.60eV~\cite{Ehrlich}. In the molecular regime, the line with solid squares is obtained according to equation ~\ref{equ:n} by using the data $\Delta E=4.2eV, R=2.393 A$, at $\zeta=1 $  for the AgBr monomer~\cite{Zhang} . From Figure~\ref{fig:figure2} we can see that at ${\zeta=1}$  , the turn-round point is anchored to $\Delta E=11.2eV $ , which corresponds to a wavelength 111nm. However, in our experiments, we observed that the turn round point was at 269nm, so the delocalization constant   must be readjusted.

\section{Effects of Delocalization}

\begin{figure}[htb]
\begin{center}
\epsfig{file=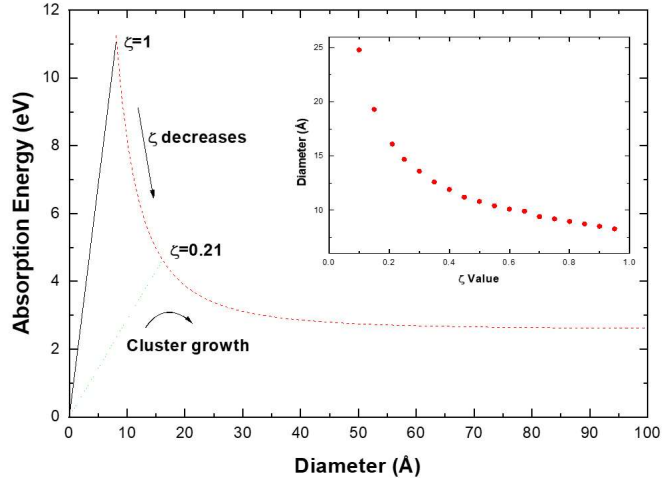,height=3in}
\caption{Absorption energy of clusters in the molecular regime as a function of the delocalization constant   and diameter of the cluster.}
\label{fig:figure4}
\end{center}
\end{figure}

      It was found that the delocalization constant affects a lot on the absorption energy. In Figure~\ref{fig:figure3}, the absorption energy are plotted versus the delocalization constant  ${\zeta}$ and the diameter R for the AgBr cluster in the molecular regime according to equation~\ref{equ:n}\index{equ:n}. We can see from Figure~\ref{fig:figure3} that for a fixed diameter, the absorption energy decreases with the decreasing of delocalization constant ${\zeta}$  . So in Figure~\ref{fig:figure2} we can fix ${\zeta}$   at a lower value, so that the turn-round point reaches the value we observed in the experiment. Figure~\ref{fig:figure4} shows the process when the delocalization constant ${\zeta}$   reduces from 1 to 0.21, the turn-round point is anchored from 11.2eV to 4.6 eV, the later, which corresponds to a wavelength 269nm, is exactly we observed in the experiment. The insert of Figure~\ref{fig:figure4} shows the diameter of cluster at the turn-round point changes with the delocalization constant ${\zeta}$  . According to Figure~\ref{fig:figure4}, in the Figure~\ref{fig:figure1}, the diameter of AgBr clusters at the turn-around point ( 269nm) is 16.2 A , in the crystal regime, the diameter of those absorbing at 273nm is 16.5A , in the molecular regime, the diameter of those absorbing at 274nm is 15.7 A. 

\section{Stability of Clusters}
Table~\ref{tab:shell} has listed the shell structure data for particles in the spherical quantum well model. When the total electron number N corresponds to a closed shell, this cluster is expected to be particularly stable. A function $\Delta _2 ( N ) $  was introduced to illustrated the special stability of closed shell cluster~\cite{Knight}:

\begin{equation}
 \ \Delta _2 ( N )= E(N+1)+E(N-1)-2E(N)
\label{equ:o}
\end{equation}

\begin{figure}[htb]
\begin{center}
\epsfig{file=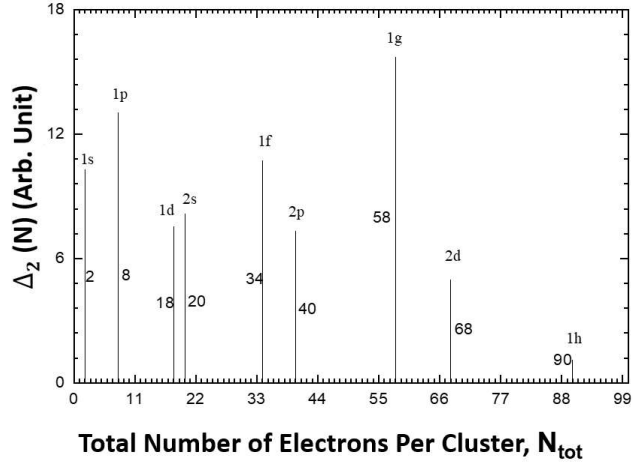,height=3in}
\caption{Stability function   changes with the total number of electrons. Only for closed shell   has relative large values.}
\label{fig:figure5}
\end{center}
\end{figure}

 $\Delta _2 ( N ) $   is related to the second derivative with respect to N of the total energy of a cluster E(N)\cite{Cohen}. The function $\Delta _2 ( N ) $   is independent of the reference energy of the free atoms, and it is a measure of the relative stability of clusters. By using the $ \chi_{nl}$  value shown in the Table~\ref{tab:shell}, a plot of $\Delta _2 ( N ) $  versus the total number of electrons per cluster (N) has been shown in Figure~\ref{fig:figure5}. From Figure~\ref{fig:figure5}, we can see that when the total electron number in a cluster is 2, 8, 18, 20, 34, 40…, etc, the $\Delta _2 ( N ) $  value is relatively large. From Table~\ref{tab:shell} we can see that all these numbers are corresponding to the electron numbers of closed shell.

\section{Conclusion}
We show that in the molecular regime, the band blue shift associated with cluster growth can be understood by a model that assume electrons are confined to a spherical potential well and the clusters are made of some basic units. A formula is given for the lowest excited electronic state energy. This expression contains an electron delocalization constant   as an adjustable parameter which, however, can be anchored to a definite value through the known transition energy at the spectra turn-around point. The stability of clusters is characterized by a function   that can be calculated by the eigenvalues of the Hamiltonian of the model.

\end{document}